# Multivariate Time-Between-Events Monitoring – An overview and some (overlooked) underlying complexities

INEZ M. ZWETSLOOT[a], TAHIR MAHMOOD[a] and WILLIAM H. WOODALL[b]

[a]Department of Systems Engineering and Engineering Management, City University of Hong Kong, Tat Chee Avenue, Kowloon, Hong Kong
[b]Department of Statistics, Virginia Tech, Blacksburg, Virginia

## Abstract
We review methods for monitoring multivariate time-between-events (TBE) data. We present some underlying complexities that have been overlooked in the literature. It is helpful to classify multivariate TBE monitoring applications into two fundamentally different scenarios. One scenario involves monitoring individual vectors of TBE data. The other involves the monitoring of several, possibly correlated, temporal point processes in which events could occur at different rates. We discuss performance measures and advise the use of time-between-signal based metrics for the design and comparison of methods. We re-evaluate an existing multivariate TBE monitoring method, offer some advice and some directions for future research.

# 1. Introduction

Monitoring methods based on time-between-events (TBE) have become increasingly popular in recent years. With TBE data, one measures the time elapsed between successive events of interest, e.g., manufacturing defects. These are also known as inter-arrival time data. Within the manufacturing realm, this framework is needed due to increasing process quality, which has resulted in so-called high-quality processes. In high-quality processes, the rate of events or defects is very low and hence the number of defects per sample over any reasonable aggregation period is usually zero or very small. For high-quality processes, time-between-events charts have been advocated as the appropriate monitoring tool (Xie et al. 2011).

A second reason for the rise of monitoring with TBE data is that it is increasingly common for measurements to be made on every item produced. This trend has been driven by the development of cheap, fast and non-destructive automated measurement devices such as sensors. With such 100 percent inspection, it is easy to obtain timestamp data of an event of interest and compute the time between consecutive events.

Monitoring with TBE data enables real-time monitoring since an observation becomes available as soon as the event happens. Whereas the alternative, monitoring counts of events over pre-specified aggregation periods, always results in a delay in monitoring until the time period, e.g., a day, is completed. Thus TBE monitoring enables real-time decision support. This can be very important in syndromic surveillance, cybercrime monitoring, warranty claims, or terrorist event detection. Some other examples of TBE data are the number of hours between failures of, for example, valves (Chen 2014), time intervals in days between explosions in coal mines (Jarrett 1979), time between outbreaks of diseases in syndromic surveillance (Sparks et al. 2019), time between communication events in a social network (Li et al. 2017), and time between accidents for monitoring occupational safety (Schuh et al. 2014).

The monitoring of TBE data for univariate data streams has a long history and goes back to the 1980's (Ali et al. 2016). More recently, some researchers have developed monitoring methods for bivariate TBE data (Xie et al. 2011; Kuvattana and Sukparungsee 2015; Sukparungsee et al. 2018) and multivariate TBE (MTBE) data (Sukparungsee et al. 2017; Khan et al. 2018; Flury and Quanglino 2018). Our aim is to discuss two complexities in the design of monitoring MTBE data and offer practical advice on how to deal with these issues appropriately.

It is helpful to consider two fundamentally different scenarios with respect to MTBE data. The first occurs when uncensored TBE vector data are observed one vector at a time. For example, consider manufactured items that can fail in several ways. If one waits until the times until each failure mode is observed for the first time for each item, then one can form a vector of uncensored failure times. We refer to this scenario as leading to *vector-based* MTBE data.

The other scenario involves the monitoring of several continuous time point processes, where each point corresponds to the occurrence of an event. The rate at which events occur could vary from process to process even when the processes are stable. An example would be the monitoring of several types of accidents in a production facility. The rates for the various types of accidents could vary with the more serious accidents occurring much less frequently than, the more minor ones. We refer to this scenario as *multivariate point process* data.

In addition to recognizing and distinguishing between these two different scenarios, it is important that the appropriate performance measures be used to evaluate methods. Due to the nature of the data, we advise to use time-between-signal based metrics for evaluation of methods, rather than run-length based metrics. The run-length typically refers to the number of points plotted on a chart until an out-of-control signal is given. This metric is not appropriate with TBE data because the times between plotted points varies. This advice is in-line with that given by Zwetsloot and Woodall (2019).

We focus on monitoring MTBE data, but it should be noted that some of the discussed issues also apply to monitoring with univariate TBE data.

## 2. The two MTBE data scenarios

### 2.1 Vector-based data

In this subsection, we illustrate that the components of MTBE vector data are not obtained simultaneously, making monitoring with complete vectors only possible after waiting for an event to occur for each of the individual processes of interest. This can result in undesirably long detection times for detecting changes in the rates.

Consider an essential part of a system, e.g., the railroad switch to guide trains from one track to another. We assume the switches have two components (A and B) that are essential and sometimes break down. The breakdowns of components A and B could be dependent. Denote by $Y$ the bivariate vector of times until the first breakdown event for each component. Often it is assumed that $Y$ follows a bivariate exponential, Weibull or gamma distribution. Figure 1 illustrates three vector possibilities corresponding to three different parts. For Part I, component A fails before component B, for part II both components fail simultaneously (highly unlikely for most applications) and for part II component A fails after component B. Monitoring cannot occur until the latter failure, resulting in a delay in incorporating information from the first failure.

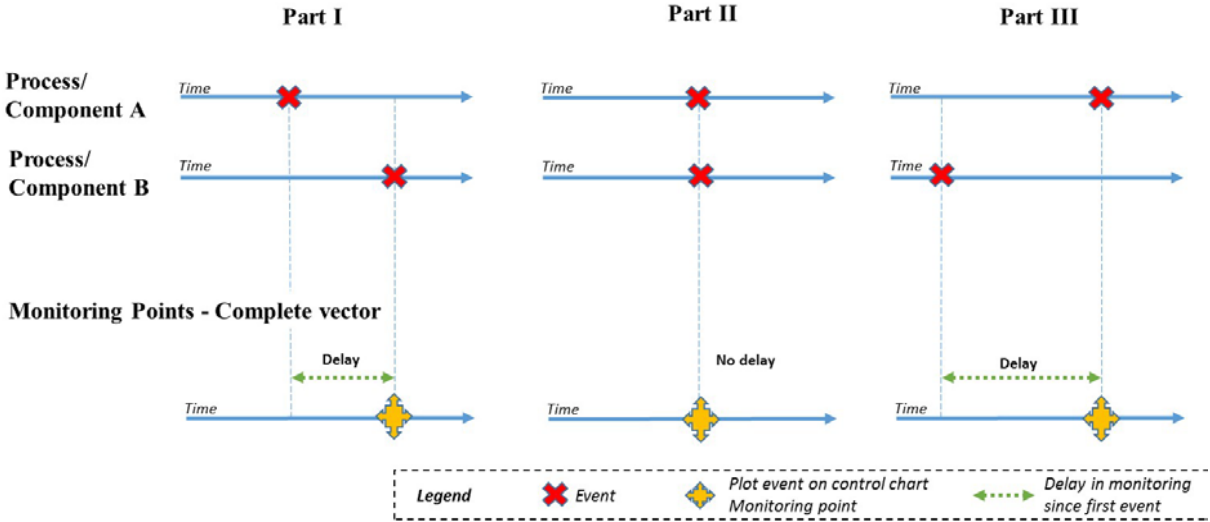

**FIGURE 1:** *Illustration of delay in monitoring when monitoring with a complete bivariate vector of TBE data.*

It is usually recommended that one monitor the times to each event using a bivariate control chart in order to detect changes in the expected times until failure. Most bivariate TBE control charts allow the plotting of the monitoring statistic only once we have observed a complete vector ***Y***. These times are indicated by the arrowed-stars in the lower part of Figure 1. This approach automatically results in a delay in incorporating the first failure time, illustrated by the dotted green arrows.

Obviously, delays are undesirable when we wish to detect changes in the process as quickly as possible. Furthermore, it is easy to see how an extension from a bivariate to a multivariate process will result in even longer delays in forming the vectors used for monitoring. All methods in the literature (see Section 4 for details) are subject to this delay in monitoring, as all current methods are based on the MTBE vector scenario.

An example of the *vector-based scenario* was described by Li et al. (2012). They considered estimation of the failure time distribution in an application of bivariate failure time data arising from the Diabetic Retinopathy Study (DRS) (Huster et al. 1989). According to Li et al. (2012), "The study was conducted by the National Eye Institute to assess the effect of laser photocoagulation in delaying the onset of severe visual loss such as blindness in the patients with diabetic retinopathy. It consisted of 197 high-risk patients. At the beginning of the study, for each patient, one eye was randomly selected for laser photocoagulation and the other was given no treatment, serving as the control. The times to blindness in both eyes were recorded in months." Perhaps unsurprisingly, Li et al. (2012) found that the times to blindness in the eyes of a patient were dependent. They also found that the failure time of the treated eye tended to be longer than the failure time for the untreated eye. A similar medical example can be found in a headache relief time study where two treatments are compared and for each patient, the relief time is recorded for each of the two treatments (Lu and Bhattacharyya 1991; Gross and Lam 1981).

Another example of *vector-based TBE data* was described by Flury and Quanglino (2018). They studied a production process, which consisted of three consecutive processes. For each batch of product, they considered the production time required for each process. Wishing to detect changes in the production times, the proposed method of Flurry and Quanglino (2018) was based on the implicit assumption that one wait until all three production processes are completed before the method can signal any change in distribution of the three production times. For a real-time solution, it would be desirable to be able to signal a change after the first or the second process have finished without waiting until we have observed an event for each of the three processes.

Another reliability related example of *vector-based TBE data* was described by Nelson (1982) and Hougaard (1989). They considered failure time of motors, where the motor only fails after three components (turn, phase and ground) each fail separately. When a part failed, it is isolated electrically and could not fail again. The motors were tested until two or three components failed.

Although all existing MTBE monitoring methods, except Li et al. (2017), are based on the vector-based approach, we think applications would be quite rare. Most situations call for the multivariate point process approach discussed in the next sub-section.

**2.2 Multivariate point process data**
To illustrate multivariate point process data, consider the application discussed by Sparks et al. (2019), who focused on syndromic disease surveillance in Australia. Using Twitter data, the authors aimed to detect specific symptoms in order to have early detection of disease outbreaks. The authors recorded the timestamp of relevant tweets and monitored the time-between-events for particular symptoms of interest. They used an exponentially weighted moving average (EWMA) based method for each type of event, i.e., a univariate approach. They illustrated their method for Fever and for Head Cold separately.

A logical next step would be to monitor these two data streams simultaneously using bivariate methods, as it can be expected that Fever and Head Cold would show dependencies. This application involves the monitoring of two point processes. Figure 2 gives an illustration of this scenario, where Process I has a naturally shorter average time-between-events compared to Process II.

Of the methods proposed in the literature only the method Li et al. (2017) is applicable for monitoring multivariate point process TBE data, although we believe this to be the much more common scenario. One approach would be to monitor each point process separately using individual control charts. The overall performance of the method, however, would depend on the correlation between the univariate processes.

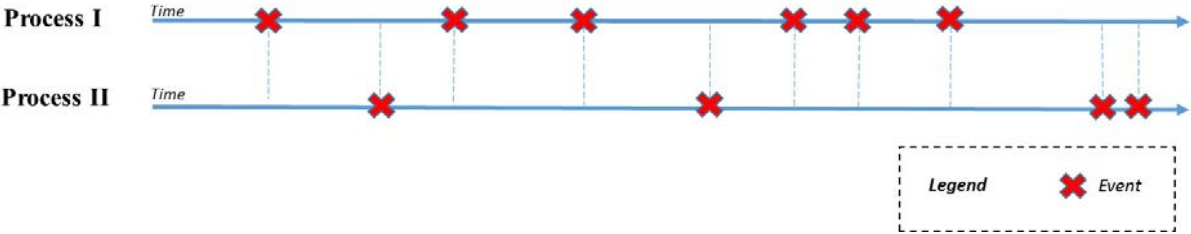

**FIGURE 2:** *Illustration of bivariate point process TBE data.*

We believe that this type of MTBE data will occur in many situations. Another example would be related to the dataset considered in Santiago and Smith (2013) and Ali and Pievatolo (2016), who were interested in the number of days in between hospital discharges of male patients. If one wished to simultaneously monitor male and female discharges, we would have a bivariate point process data stream.

Two medical related examples of *point process MBTE data* are given by Byar (1980) and Chiou et al. (2018). Byar (1980) studied bladder cancer recurrence data and they collected the time between recurrences for multiple patients. The recurrence of bladder cancer for some patients happens more frequently than for others. Chiou et al. (2018) studied recurrence of skin cancer and also here frequency of recurrences differs between patients as well as between gender.

## 3. Performance metrics: use ATS not ARL
In this section, we provide our perspective and advice on the use of performance metrics to evaluate the statistical performance of MTBE monitoring methods. This perspective was also given by Zwetsloot and Woodall (2019).

Currently, performance metrics used for MTBE methods have been based on run lengths, i.e., the number of plotted statistics until an out-of-control signal is observed. However, it is far more appropriate to use time-to-signal metrics because the time between plotted statistics varies when monitoring with TBE data. As the objective of monitoring TBE data is to detect process changes as soon as possible, the average time until such a change is detected should be used as a performance metric.

Defining time-to-signal metrics can be tricky for MTBE data. The definition may depend on the application under consideration. Here we give possible definitions for the Average Time-to-Signal (ATS) for vector-based data and for multivariate point process data.

**3.1 ATS for vector-based data**
For our purposes, we denote the multivariate time-between-events recorded for a $p$-variate vector-based MTBE data by $\boldsymbol{Y_i} = \{Y_{i1}, Y_{i2}, \ldots, Y_{ip}\}$ for $i = 1, 2, 3, \ldots$, with $i$ indicating the person or item observed. Here the components of the vector denote the time until an event occurred. For example, we might have the times since treatment until the left eye ($Y_{i1}$) and the right eye ($Y_{i2}$) went blind for person $i$ in the diabetic retinopathy study or the times passed since a motor was started until component 1, 2 and 3 broke down.

Next, we assume a probability distribution for the components of the vectors $\boldsymbol{Y_i}$ which has a parameter vector $\boldsymbol{\theta_i}$. We are interested in monitoring the process to detect a change in the parameter vector from $\boldsymbol{\theta_0}$ to $\boldsymbol{\theta_1}$ as quickly as possible. Assume we observe $v$ products or persons from the in-control state. Therefore, for $i \leq v$ we have $\boldsymbol{\theta_i} = \boldsymbol{\theta_0}$ and for $i > v$ we have $\boldsymbol{\theta_i} = \boldsymbol{\theta_1}$. Next, for each of the event time vectors, we define $T_i = \max_{j=1,2,\ldots,p} \boldsymbol{Y_i}$, so now $T_i$ corresponds to the waiting time until we observe the complete vector $\boldsymbol{Y_i}$. Note that $T_i$ is a univariate data stream, denoting the time until each vector is complete.

With a control chart approach, an alarm is observed at time $t_A$ when the monitoring statistics $M_{i_A}$ (e.g., a Hotelling's T-squared, or a multivariate EWMA, or a multivariate cumulative sum (CUSUM) statistic) say the $i_A$-th event exceeds the pre-specified threshold. Note that the alarm time can be computed as $t_A = \sum_{i=1}^{i_A} T_i$. Figure 3 illustrates the time to signal $t_A$ for a bivariate TBE process where we assume that we observe an out-of-control signal at the fourth vector observation, $i_A = 4$ which leads to an event-time of $t_A = 15$.

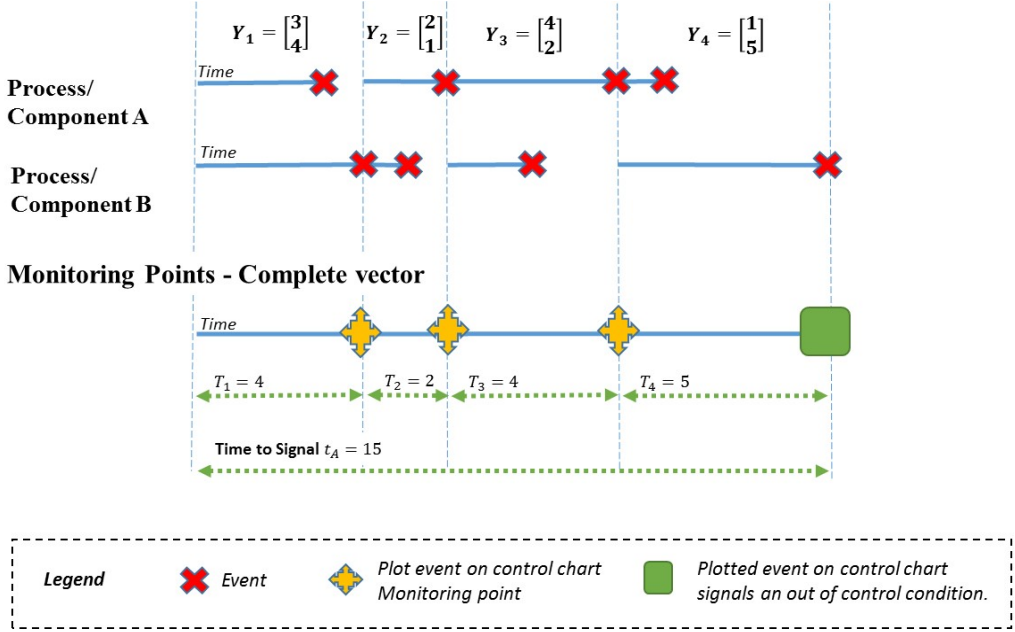

**Figure 3:** Illustration of how to compute time-to-signal metrics for vector-based MTBE data.

The in-control ATS can now be defined as $ATS_0 = E[t_A | v = \infty]$. The in-control ATS is defined as the expected time until a signal is observed given no change in the process. Here $v = \infty$ indicates no change has occurred in $\theta_i$. On the contrary, the in-control average run length is defined as $ARL_0 = E[i_A | v = \infty]$, where $i_A$ is the sample number corresponding to the signal time $t_A$, i.e., the expected number of events until a signal is observed. In our comparison study in Section 5, we will employ the average time-to-signal (ATS), as defined above, as the performance measure. Li et al. (2014) provided an overview of computation methods for ARL and ATS metrics, but their discussion mostly focuses on the ARL.

### 3.2 ATS for multivariate point process data

For our purpose of defining the ATS for multivariate point processes, we denote the multivariate time-between-events recorded for a *p*-variate point process by $\boldsymbol{Y_j} = \{Y_{1j}, Y_{2j}, Y_{3j} \ldots\}$ for $j = 1, 2, \ldots, p$. Here the components of the vector represent the time passed since the previous event or the time passed since the start of monitoring (for the first component). We assume a probability distribution for the components of the vectors $\boldsymbol{Y_j}$ which have a parameter vector $\boldsymbol{\theta_j}$. We are interested in monitoring the process to detect a change in the parameter vector from $\boldsymbol{\theta_0}$ to $\boldsymbol{\theta_1}$ as quickly as possible. Assume that the process is in-control until time $\tau$ and moves to an out-of-control state after time $\tau$. When $t \leq \tau$ we have $\boldsymbol{\theta_j} = \boldsymbol{\theta_0}$ and for $t > \tau$ we have $\boldsymbol{\theta_j} = \boldsymbol{\theta_1}$. Next, for each of the *p*-event time vectors define a vector $\boldsymbol{T_j} = \{T_{1j}, T_{2j}, T_{3j}, \ldots\}$ corresponding to the waiting time until each event occurred, where $T_{kj} = \sum_{i=1}^{k} Y_{ij}$. With a Shewhart control chart approach, an alarm is observed at time $t_A$ when an observed time-between-event is either smaller or larger than the pre-specified thresholds, i.e., $Y_{ji} > h_{jU}$ or $Y_{ji} < h_{jL}$ and $t_A = T_{ji}$. Of course, an EWMA or CUSUM method could be used instead.

The in-control ATS can now be defined as $ATS_0 = E[t_A | \tau = \infty]$. The in-control ATS is defined as the expected time until a signal is observed given no change in the process. Here $\tau = \infty$ indicates no change has occurred in $\boldsymbol{\theta}$. Note that due to the varying rates of the point processes it is difficult to define $i_A$, the expected number of events until a signal is observed. In the out-of-control case, we are interested in the time $t_A$-$\tau$.

### 3.3 Zero-state ATS
One measure of detection ability is the ATS given that the change occurs immediately at the beginning of Phase II monitoring ($\tau = 0$). This so-called zero-state performance is widely used; however, it only shows a partial picture of the performance.

### 3.4 Steady-state ATS
Steady-state performance considers the time from a change to a signal when the process changes at any random time point after monitoring begins in Phase II, i.e., $v > 0$ or $\tau > 0$. We consider steady-state performance to be much more realistic than zero-state performance. See Frisén et al. (2010) for a detailed discussion on zero-state vs steady-state performance.

## 4. Literature review and description of methods

In recent years, several researchers have developed methods for monitoring MTBE vector-based data. In this section, we review this literature and describe a few selected monitoring methods in detail. We focus our literature review on multivariate control charts that are suitable for monitoring multivariate time-between-events data. Note that, we excluded one article (Mukherjee at al. 2018) because its data structure assumed that the complete vectors become available at a single time point which is not a TBE vector-based data structure.

Various authors have reviewed multivariate statistical process control methods (Bersimis et al. 2007; Woodall and Montgomery 2014; Dhini and Surjandari 2016; Bersimis et al. 2018). However, none of these authors reviewed or considered multivariate time-between-events monitoring. Ali et al. (2016) gave an overview of control charts for high-quality processes. Their overview focused primarily on univariate methods. They advocated the use of time-between-events control charts to overcome difficulties in monitoring high-quality processes.

We first review methods for vector-based MTBE data in Section 4.1. Followed by a review on methods for point process data in Section 4.2.

## 4.1 Methods for monitoring vector-based MTBE data

All methods, except Li et al. (2017), are designed for monitoring vector-based data. With these methods one plots the monitoring statistic when the observed vector is complete. Hence, each of these methods will have a built-in delay in detecting process changes.

### 4.1.1 MEWMA for multivariate TBE data

The first paper considering bivariate TBE monitoring was Xie et al. (2011). The authors proposed a multivariate exponentially weighted moving average (MEWMA) control chart for vectors of exponentially distributed TBE data. They compared their method to the use of paired univariate *t*-charts and paired univariate EWMA charts. Since this method is the most cited and the original method, we describe it here in some detail and include it in our comparison study in Section 5.

Xie et al. (2011) proposed applying the MEWMA control chart of Lowry et al. (1992) to bivariate exponential data as well as to transformed data. We consider the untransformed version as their comparison does not show much difference in performance. Xie et al. (2011) considered bivariate lifetime vectors $\boldsymbol{Y}_i = (Y_{i1}, Y_{i2})$ from Gumbel's bivariate lifetime model (Gumbel 1960). Their monitoring statistics are

$$E_i^2 = \frac{2-\lambda}{\lambda} \boldsymbol{z}_i^T \boldsymbol{\Sigma}_Y^{-1} \boldsymbol{z}_i, \qquad (1)$$

where $\boldsymbol{z}_i = \lambda(\boldsymbol{Y}_i - \boldsymbol{\mu}_Y) + (1-\lambda)\boldsymbol{z}_{i-1}$ for $i = 1, 2, \dots$, $0 < \lambda \leq 1$ is the EWMA smoothing constant, and $\boldsymbol{\mu}_Y$ and $\boldsymbol{\Sigma}_Y$ are the assumed known in-control process mean vector and covariance matrix of $\boldsymbol{Y}$. It is noted that the initial value of the MEWMA statistic ($\boldsymbol{z}_0$) is set at the zero vector. The process is considered to be out-of-control when $E_i^2$ exceeds the control limit $h > 0$. Xie et al. (2011) obtained $h$ using a simulation procedure by setting the in-control ARL to a pre-specified value.

Another MEWMA chart for MTBE data was proposed by Khan et al. (2018). They first apply the EWMA to the univariate data and then combine these statistics into one monitoring statistic. Also, Flury and Quaglino (2018) proposed a MEWMA chart for gamma distributed MTBE data and applied it to an example of production time data.

Xie et al. (2011), as well as Khan et al. (2018) and Flury and Quaglino (2018), evaluated their methods based on ARL and therefore their results on the performance of the proposed charts are difficult to interpret.

**4.1.2 Copula based charts for MTBE monitoring**

Recently, some TBE monitoring methods have been proposed based on copula modelling. For multivariate data, it is often difficult to obtain a joint density function that reflects the dependency structure precisely. One possible solution is to use copula modelling; this approach allows one to model data using marginal distributions and a separate copula for the dependency structure.

For bivariate TBE processes, Kuvattana and Sukparungsee (2015) proposed a MEWMA, multivariate double EWMA and multivariate CUSUM charts when the data are modelled using copulas. The multivariate CUSUM chart is designed to detect increases in the mean vector of the underlying data only. Furthermore, Sukparungsee et al. (2017) proposed a monitoring method based on copula modelled data for the monitoring of trivariate TBE data. Sukparungsee et al. (2018) used copulas for modelling dependencies and implemented this idea for a Hotelling's $T^2$ chart. All these authors compared and evaluated their proposed methods using the ARL, hence their results are difficult to interpret.

**4.2 Methods for monitoring multivariate point process data**

Li et al. (2017) proposed a change-point detection method for multivariate point process event data. Their method is specifically designed for social network data. They use a sequential hypothesis test and derive the likelihood ratios for the point processes and demonstrate their methods on a network dataset from Twitter and Memetracker. Like all other reviewed papers, Li et al. (2017) use ARL based metrics to evaluate statistical performance. Li et al. (2017) is the only paper that is designed to monitor multivariate point process data.

## 5. Evaluation of existing methods using ATS

All reviewed control charts were evaluated using ARL metrics. As this metric does not reflect the true performance of the methods (as argued in Section 3), we re-evaluated the widely cited MEWMA method proposed by Xie et al. (2011) and a paired univariate EWMA approach (PEWMA) using the ATS metric as defined in Section 4.1.

The data model we used for generating vector-based bivariate TBE data was Gumbel's bivariate exponential distribution (Gumbel 1960), with survival function

$$S(y_1, y_2) = \exp-\left[\left(\frac{y_1}{\theta_{01}}\right)^{\frac{1}{\delta}} + \left(\frac{y_2}{\theta_{02}}\right)^{\frac{1}{\delta}}\right]^{\delta}. \qquad (2)$$

Here $\theta_{01}$ and $\theta_{02}$ are the scale parameters and $\delta$ is the dependence parameter. The value $\delta = 1$ corresponds to independence.

We compared the MEWMA chart as proposed by Xie et al. (2011) and defined in equation (1) to a pair of univariate EWMA charts. As univariate charts do not experience a delay in monitoring due to waiting for a complete vector to form, we included this paired method in our comparison. We used a pair of one sided univariate EWMA charts by following Gan (1998). We used upper-sided statistics when we expect increases in the process parameters leading to an increase in the waiting times. We have

$$z_{i1} = \max(\theta_{01}, \lambda y_{i1} + (1-\lambda)z_{i-1\ 1}) \text{ and}$$
$$z_{i2} = \max(\theta_{02}, \lambda y_{i2} + (1-\lambda)z_{i-1\ 2}) \quad (3)$$

where $z_0 = (\theta_{01}, \theta_{02})$. The first chart signals when the statistic $z_{i1}$ exceeds the upper control limit $U_1$. The second chart signals when the statistic $z_{i2}$ exceeds its limit $U_2$.

We use two lower sided EWMA statistics when we expect *decreases* in the process parameters:

$$z_{i1} = \min(\theta_{01}, \lambda y_{i1} + (1-\lambda)z_{i-1\ 1}) \text{ and}$$
$$z_{i2} = \min(\theta_{02}, \lambda y_{i2} + (1-\lambda)z_{i-1\ 2}), \quad (4)$$

where $z_0 = (\theta_{01}, \theta_{02})$. The first chart signals when the statistic $z_{i1}$ falls below the lower control limit $L_1$. The second chart signals when the statistic $z_{i2}$ falls below its limit $L_2$.

We used steady-state ATS as the performance measure and evaluated the ATS using 100,000 Monte Carlo simulations. For the steady-state setup, we assume that the first 50 samples are generated from the in-control model and shifts are introduced from the 51[th] sample. We considered four different in-control scenarios using the survival function as defined in Equation (2). Model 1 and Model 2 have $\theta_{01} = 1$ and $\theta_{02} = 2$ representing the case that the mean event time is almost similar for both sub-processes. For Models 3 and 4 we have $\theta_{01} = 10$ and $\theta_{02} = 2$ representing the case that the mean event time is considerably different for the sub-processes. Models 1 and 3 are based on independence ($\delta = 1$) and Models 2 and 4 represent dependent data ($\delta = 0.5$).

We obtained the values of the control limits through simulation by setting the in-control ATS equal to 200. We considered six out-of-control scenarios where $\theta_{01}$ and/or $\theta_{02}$ either decrease by 50% or increase by doubling. ATS results are displayed in Table 1 and Figure 4.

| In-control models | Model 1 $\theta_{01}=1, \theta_{02}=2, \delta=1$ | | Model 2 $\theta_{01}=1, \theta_{02}=2, \delta=0.5$ | | Model 3 $\theta_{01}=10, \theta_{02}=2, \delta=1$ | | Model 4 $\theta_{01}=10, \theta_{02}=2, \delta=0.5$ | |
|---|---|---|---|---|---|---|---|---|
|  | MEWMA | PEWMA | MEWMA | PEWMA | MEWMA | PEWMA | MEWMA | PEWMA |
| In-control | 200 | 200 | 201 | 202 | 200 | 202 | 200 | 200 |
| Control limits | $h=6.90$ | $L_1=0.5685$ $L_2=1.15$ | $h=7.40$ | $L_1=0.578$ $L_2=1.15$ | $h=3.49$ | $L_1=6.685$ $L_2=1.33$ | $h=3.46$ | $L_1=6.86$ $L_2=1.37$ |
| Out-of-control Low-Low: $\theta_{01}^*=0.5\theta_{01}$, $\theta_{02}^*=\theta_{02}$ | 57.37 | **23.49** | 42.09 | **21.50** | 51.99 | **31.42** | 40.57 | **23.03** |
| $\theta_{01}^*=\theta_{01}$, $\theta_{02}^*=0.5\theta_{02}$ | 40.98 | **16.58** | 27.08 | **14.16** | 94.52 | **52.10** | 79.05 | **42.07** |
| $\theta_{01}^*=0.5\theta_{01}$, $\theta_{02}^*=0.5\theta_{02}$ | 18.15 | **9.28** | 34.12 | **9.96** | 37.19 | **20.13** | 49.41 | **21.90** |
| Control limits | $h=6.90$ | $U_1=1.63$ $U_2=3.255$ | $h=7.40$ | $U_1=1.63$ $U_2=3.26$ | $h=3.49$ | $U_1=14.1$ $U_2=2.82$ | $h=3.46$ | $U_1=13.85$ $U_2=2.79$ |
| Out-of-control Up-Up: $\theta_{01}^*=2\theta_{01}$, $\theta_{02}^*=\theta_{02}$ | 25.57 | **24.85** | **19.41** | 22.04 | 105.53 | **89.20** | 90.72 | **88.05** |
| $\theta_{01}^*=\theta_{01}$, $\theta_{02}^*=2\theta_{02}$ | 35.79 | **35.07** | **30.41** | 34.91 | 58.10 | **47.37** | 46.70 | **40.10** |
| $\theta_{01}^*=2\theta_{01}$, $\theta_{02}^*=2\theta_{02}$ | 24.26 | **23.33** | 28.27 | **24.47** | 71.52 | **53.63** | 82.67 | **52.30** |

**Table 1:** *ATS values for the MEWMA control chart and the paired univariate EWMA control charts. Four in-control models are considered with six out-of-control scenarios each. In-control ATS value is designed to be 200. Boldface indicates lowest ATS for that in-control model and out-of-control scenario.*

We see that MEWMA and PEWMA methods each can have the shortest out-of-control ATS depending on the scenario considered. In Figure 4, we see that the MEWMA chart and PEWMA charts take approximately equally long on average to detect most changes. This result is surprising as one would expect that the multivariate method like the MEWMA would outperform two univariate methods, especially when the data are dependent (Model 2 and Model 4). Often the PEWMA chart is even quicker in detecting changes (more dots above the diagonal line in Figure 4). This is because the use of the MEWMA chart requires waiting until full vectors of observations are available. The PEWMA chart, however, can signal a change in a sub-process without waiting for events from the other sub-process to be observed. From this result, we can conclude that the current multivariate method for vector-based TBE data does not outperform the use of paired univariate methods.

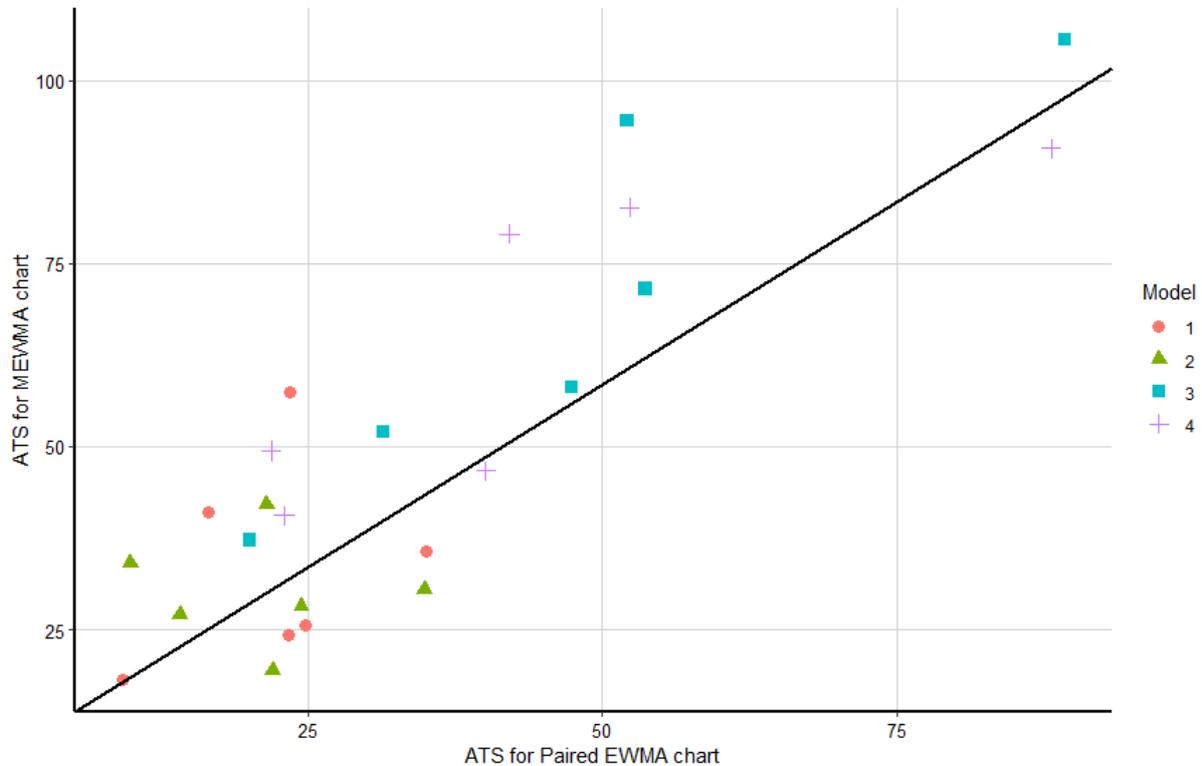

**Figure 4:** *ATS values for the MEWMA control chart and the paired univariate EWMA control charts. Four in-control models are considered with six out-of-control scenarios each. In-control ATS value is designed to be 200.*

## 6. Other issues
In this section, we discuss some other issues related to monitoring multivariate TBE data.

### 6.1 Model selection for vector-based TBE data
Prospective monitoring is typically called Phase II monitoring. This relies on the process knowledge gained in the Phase I analysis of historical data. Jones-Farmer et al. (2014) gave an overview of Phase I objectives and methods. One aspect of Phase I is to determine an appropriate distributional model for the data and to estimate the unknown parameters. Here we discuss some considerations regarding the appropriate choice of a distribution function for vector-based TBE data.

Models for bivariate failure-time data go back to the 1960's. Many different models are available in the literature (Kotz et al. 2000). One important consideration to keep in mind during the selection process of the appropriate model is the failure mechanism of the underlying process. Usually, the following classes of models are considered:

- **Class 1:** These models consider a random external stress factor influencing both components. The Gumbel's bivariate exponential (Gumbel 1960) and Hougaard's bivariate Weibull (Hougaard 1986) models are the examples in this class.

- **Class 2:** The failure-times in these models result from internal events of the system. For example, we might have a system where the failure of one component induces an additional burden on the remaining one (e.g., kidneys) or, alternatively, the failure of one may relieve somewhat the burden on the other (e.g., competing species). The Freund (1961) and Block and Basu (1974) models are examples of this class.

- **Class 3:** In these models, failure of the components occurs due to a third source common shock. Examples are the Marshall–Olkin (Marshall and Olkin 1967), Downton (Downton 1970), Proschan and Sullo (1974) and Sarkar (Sarkar 1987) models.

- **Class 4:** Models in this class combine characteristics of both class 2 and 3. The Friday and Patil (1977) and the Tosch and Holmes (1980) models lie in this category.

- **Class 5:** In this class, a system having two components can be classified into three types of functioning states: normal, unsatisfactory and failed. See, for example, the Raftery model (Raftery 1984).

There exist more classes of bivariate exponential distributions. For more details see Balakrishnan and Lai (2009). All these models are bivariate data models for vector-based TBE data.

### 6.2 Monitoring counts or Bernoulli events

An alternative approach to monitoring time-between events is the use of counts of events over a certain aggregation time period or considering Bernoulli variables representing the occurrence or non-occurrence of the event. See Szarka and Woodall (2011) for a review of methods for monitoring with Bernoulli data.

There is a large literature on (multivariate) count data monitoring. For overviews see Saghir and Lin (2015) and Mahmood and Xie (2019). Note that these methods always result in a delay in detection as counts need to be accumulated over a certain fixed time interval, e.g., each day or each hour. Consequently, one has to wait until the end of each day or hour, as the case might be, to be able to signal a change in the process. These issues were discussed by Schuh et al. (2013) and reviewed by Zwetsloot and Woodall (2019).

## 7. Conclusions and research directions

We have reviewed the literature for multivariate time-between-events monitoring. We have identified two scenarios for data collection when designing TBE control charts. The point process application is much more common than the vector-based scenario. We have given some advice regarding the use of performance metrics and have shown that the choice of performance metric is important. We advise use of the steady-state time-to-signal based performance metrics when evaluating and comparing monitoring methods for TBE data.

Given the development of real-time data acquisition, we predict that MTBE monitoring will become more popular. Further research is necessary for the development of methods. Some topics for future research include the following:

- Nearly all research work thus far has focused on monitoring with vector-based data. Thus, monitoring methods need to be developed for multivariate point processes. One possible research direction is monitoring the superimposed process, i.e., where multiple data streams are combined into one data stream. For example, an event is defined to occur when an event occurs in any of the sub-processes. Some initial work in this direction was presented by Sparks (2019). Note that this approach may work well if several processes have increases in rates. It would be expected to work poorly in detecting increased rates for relatively rare events.
- All the work reviewed is designed to detect sustained shifts in the underlying process. The detection of transient shifts in rates has yet to be studied.
- Nearly all research work thus far, for vector-based data, allows monitoring only when the complete data vectors are observed. This prolongs the detection time until an event is observed in each of the sub-processes. Using univariate charts for each sub-process meets this objective; however, the underlying correlation structure between the processes will then tend to be ignored. Multivariate methods need to be developed that enable monitoring as data become available.
- Most research has focused on the monitoring performance in Phase II, the prospective monitoring phase. We have found no work that focussed on methods for the Phase I analysis.
- Non-parametric schemes for MTBE monitoring could be useful.
- It is well known that parameter estimation severely influences the performance of control charts. Estimation of MTBE models is not an easy task. The effect of estimation error and model misspecification needs to be studied for the MTBE control charts.
- One additional issue for TBE data is visualization. Traditionally TBE data has been plotted on control charts just like any other process. The Minitab standard for TBE data (the t-chart) shows the time on the y-axis and the event number on the x-axis. This way of visualization turns around our natural way of thinking because we are used to having time on the x-axis. Alternative visualization techniques would benefit the user.
- No monitoring methods exist for MTBE data when auxiliary information is available.

Finally, we encourage authors to justify the selected performance metric used, as this has a major influence on performance comparisons. For TBE data time-between-signal metrics are preferred.